\documentclass[aps,prd,reprint]{revtex4-2}
\usepackage[breaklinks,colorlinks=true]{hyperref}
\usepackage{xcolor}
\usepackage{bm}
\usepackage{amsmath}
\usepackage{graphicx}
\usepackage{subcaption}

\newcommand{\unit}[1]{\,\mathrm{#1}}

\begin{document}

\title{Evolution of chirality in a multiphoton pair production process}
\author{Chengpeng Yu}
\email{yu.chengpeng@nt.phys.s.u-tokyo.ac.jp}
\affiliation{Department of Physics, The University of Tokyo, 7-3-1 Hongo, Bunkyo-ku, Tokyo 113-0033, Japan}

\begin{abstract}
Recent years, multiphoton pair production has become one of the most promising approaches to investigate the Schwinger effect. However, the production and evolution of chirality, a key topic in the study of this effect, has not been thoroughly considered in the context of multiphoton pair production. In this work, as the first step of filling this gap, we used the Dirac-Heisenberg-Wigner formalism to study the production and evolution of chirality in vacuum under the excitation of the spatially homogeneous electric and magnetic fields $\bm{E}(t)$ and $\bm{B}(t)$ that satisfy $\bm{E}(t)\parallel\bm{B}(t)$ and are only nonzero in a short time span $0<t<\tau$, which serve as a simplified model of the laser beams in multiphoton pair production experiments. Based on analytical calculation, we discovered that, after the external fields vanish, an oscillation of pseudoscalar condensate occurs in the system, which leads to the suppression of the chirality of the produced fermion pairs; at the same time, it introduces a special fermion energy $\epsilon_p=\sqrt{3} m$ at which the chiral charge distribution of the fermions maximizes. This novel phenomenon could help us identify different types of products in future multiphoton pair production experiments.
\end{abstract}

\maketitle

\section{Introduction}

In modern physics, quantum electrodynamics (QED) stands as one of the most precise theories. Nonetheless, numerous nonlinear aspects of the theory still remain untested. Among these, one profound phenomenon is the Schwinger effect. Discovered by J. Schwinger at 1951 and also discussed by several earlier scholars \cite{Schwinger1951,Sauter1931,Heisenberg1936}, this effect shows that, in the presence of an external electric field, the vacuum in QED becomes unstable, leading to the production of fermion-antifermion pairs \cite{Hebenstreit2011}. The Schwinger effect is captivating because of its nonperturbative nature, which arises since the coupling constant times the external field strength becomes so large that the vacuum at infinite past and infinite future becomes significantly different. In this way, the effect exhibits the nontrivial properties of quantum vacua \cite{Hebenstreit2010,Copinger2020}, providing valuable insights into the mystery of the chiral magnetic effect \cite{Fukushima2008,Copinger2018,Fukushima2023} and the Floquet vacuum engineering \cite{Yamada2021,Fukushima2023}, among others. In addition to its theoretical importance, this effect also plays a key role in high-energy heavy-ion collisions, particularly ultraperipheral collisions \cite{Adam2021,Gould2019,Moreau2017,Hattori2017}. Hence, pursuing the direct measurements of Schwinger effect is of crucial importance.

Despite this, due to the high field strength threshold ($E_{cr} = m^2c^3/(e\hbar)=1.3\times10^{16}\unit{V/cm}$), measuring the Schwinger effect in a pure gauge field setup still remains a challenge \cite{Hu2023,Kohlfurst2022}. (By ``pure,'' we mean that contrary to the ultraperipheral collisions, etc., there are only photons without other particles in the system.) To tackle this problem, special approaches need to be applied, and one possibility is multiphoton pair production. The basic idea is that although the field strength threshold is still a few orders of magnitude out of the reach of the current facilities \cite{Hu2023,Yoon2021}, dynamic fields can drastically reduce the required field strength \cite{Ababekri2019,Taya2020,Schutzhold2008,Dunne2009}. Hence, the collision between strong laser beams and the resulting pair production process that involves more than two photons becomes an effective way to study the Schwinger effect \cite{Hu2010,Esnault2021,Dai2021}. This idea was tested at the beginning of this century and a few pairs are observed \cite{Burke1997,Bamber1999}. Recent years, with the development of high intensity laser technology, multiphoton pair production has received more and more attention \cite{Kohlfurst2022,Ababekri2019,Hu2023}, and current study shows that apart from the field strength, different degrees of the freedom of photons, such as spin \cite{Kohlfurst2022}, pulse length \cite{Hu2023}, and spatial profiles \cite{Hebenstreit2011}, could influence the vacuum in different interesting ways. Thus, compared with the static field Schwinger effect, the process with dynamic photons exhibits additional information, which allows us to investigate the nonlinear regime of QED from a wider variety of aspects.

In addition to multiphoton pair production, the production and evolution of chirality is another widely discussed topic in the context of Schwinger effect. Chirality, or chiral charge, is defined as zeroth component of the fermion axial current $j_5^\mu = \bar{\psi} \gamma^\mu \gamma_5 \psi$ ($\psi$ is the Dirac field, $\gamma^\mu$ and $\gamma_5$ are the Dirac matrices with $\mu=0,1,2,3$), and can be intuitively understood as the number density difference between the right-handed and left-handed fermions. According to the famous axial Ward identity $\partial_\mu j_5^\mu = e^2/(2\pi^2) \bm{E}\cdot\bm{B} + 2m \bar{\psi} i\gamma_5 \psi$ ($\bm{E}$ and $\bm{B}$ are the electric and magnetic field strength, $e$ is the coupling constant, and $m$ is the mass of the fermions), if the electromagnetic field that induces the Schwinger effect satisfies $\bm{E}\cdot\bm{B}\neq 0$, the fermion-antifermion pairs emerging from vacuum would have nonzero chirality. This, in turn, would induce the chiral magnetic effect, which is a macroscopic quantum transport phenomenon that allows the excitation of electric current by magnetic field \cite{Fukushima2020,Kharzeev2016}. In recent years, the observation of the chiral magnetic effect in high-energy systems has been a hot topic \cite{Abdallah2022,Shovkovy2022,Abdulhamid2023}. Thus, chirality production via Schwinger effect has attracted much attention of the community. The discussions about the resulting pseudoscalar condensate and chiral chemical potential \cite{Fang2017}, the difference between equilibrium and out-of-equilibrium observables \cite{Copinger2018,Copinger2020}, the influence of magnetic helicity \cite{Aoi2021}, the regularization schemes \cite{Aoi2021,Fukushima2023}, and the worldline formalism treatment beyond the constant background fields \cite{Schubert2023}, to name but a few, are actively underway. These investigations may provide us with a chance to observe chiral magnetic effect in vacuum,  which would lead to more clear signal than those from heavy ion collisions \cite{Abdallah2022}.

Despite the importance of chirality production and evolution, to the best of our knowledge, they have never been thoroughly investigated in the multiphoton pair production setup. So far, when discussing chirality production, the most frequently assumed external fields are static fields \cite{Copinger2018,Copinger2020,Fang2017}. Even in the few cases when dynamic fields are considered, published works often assumes fixed $\bm{B}$ \cite{Aoi2021}, or simply uses the perturbative approach \cite{Taya2020}. However, chirality production in multiphoton pair production is both possible and important. As an argument of possibility, the $\bm{E}\cdot\bm{B}\neq0$ field configurations can be easily realized in laser based experiments, for example, by colliding two laser beams with different polarization together, which will be discussed in detail in Sec. \ref{sec:sys-conf} \cite{Taya2020}. As an argument of importance, multiphoton pair production is one of the most promising approaches to observe Schwinger effect, so it would be interesting to consider chirality production, which is an important topic of Schwinger effect, in this setup. Moreover, in multiphoton pair production, both the electric and magnetic fields are short pulses, so the magnetic helicity $H^{M} = \int d^3\bm{x} \bm{A} \cdot \bm{B}$ ($\bm{x}$ is the spatial coordinate, $\bm{A}$ is the vector potential) will vanish in the infinite past and future, which is considerably different from the usual case where the magnetic fields are constant such that $H^M$ does not vanish in the future \cite{Aoi2021}. To account for the difference, a new theoretical analysis is necessary.

Considering the above possibility and importance, in this paper, as the first step towards understanding the chirality production and evolution in multiphoton pair production, we performed analytical computation to predict the excitation of fermion pairs by electric and magnetic pulses with $\bm{E}\parallel\bm{B}$, as well as the evolution of the pairs after the excitation. We derived the chiral charge distribution at different time $t$ and discussed the characteristics of this distribution that could be observable in future experiments and decipher interesting information about the process.

The paper is organized as follows: Sec. \ref{sec:sys-conf} introduces the external field that induces multiphoton pair production; Sec. \ref{sec:DHW} reviews the Dirac-Heisenberg-Wigner (DWH) formalism, which is the tool to analyze the evolution of the system; Sec. \ref{sec:gen-sol} derives the formal solution of the DHW equation of motion which we would like to use for latter computation; Sec. \ref{sec:ex-stage} analyzes the evolution of this system before the vanishing of the external fields; Sec. \ref{sec:free-stage} analyzes the evolution afterwards; Sec. \ref{sec:result} presents and discusses the calculation results; finally, Sec. \ref{sec:concl} summarizes the findings and presents future perspectives.

\section{The External Fields\protect\label{sec:sys-conf}}

To start with, we follow the suggestion of \cite{Taya2020} and consider two counter-propagating laser beams, characterized by the following vector potentials in the Coulomb gauge: $\bm{A}_1= A \sin(kx-kt) \bm{e}_1$, $\bm{A}_2= A \sin(-kx-kt) \bm{e}_2$, with $\bm{e}_1=(1,0,0)^T$, $\bm{e}_2=(\cos\phi,\sin\phi,0)^T$ as the polarization vectors, $\phi$ as an arbitrary angle, $k$ as the photon momentum, and $A$ as the field amplitude.

From this setup, we can show that at the $kx\ll 1,kt\ll 1$ space-time region, where the pair production occurs, the magnetic field and electric field satisfies $\bm{B}=\cos\phi/(1+\sin\phi) \bm{E}$. Hence, we obtain a $\bm{E}\parallel\bm{B}$ field with the ratio $|\bm{E}|/|\bm{B}|$ depends on $\phi$, with which we can study chirality production.

Also, we need to take into account the finite length of the beams, so the $\bm{E}$ and $\bm{B}$ discussed above should be $\bm{E}(t)$ and $\bm{B}(t)$ which is only nonzero in the time span $0<t<\tau$, with $\tau$ as a small time value that characterizes the length of the beam.

Finally, for latter convenience, we rotate the $yOz$ plan such that the electric field points at the $z$ direction, then the time-dependent electric and magnetic fields becomes
\begin{align}
    \bm{E}(t) = E(t) \bm{e}_z,\label{eq:bkgr_E}\\
    \bm{B}(t) = B(t) \bm{e}_z,\label{eq:bkgr_B}
\end{align}
with $\bm{e}_z$ is the $z$-oriented unit vector. This is the external field we would use for latter computation. Furthermore, we define $E_0 = \frac{1}{\tau} \int_0^\tau E(t) dt$, $B_0 = \frac{1}{\tau} \int_0^\tau B(t) dt$.

\section{Review of Dirac-Heisenberg-Wigner Formalism\protect\label{sec:DHW}}

The pair production under strong background fields can be studied either by the DHW formalism \cite{Bialynicki-Birula1991} or the worldline formalism \cite{Schubert2023,Schmidt1993,Esposti2023,Amat2022}. As the DHW formalism is particularly suitable for studying the real-time evolution of the system induced by time-dependent fields, in this paper we would like to use this approach \cite{Ababekri2019}. Hence, the basic aspect of the DHW formalism would be reviewed in this section.

With the DHW formalism, the fermions produced in the system is described by the gauge-covariant Wigner function, defined as
\begin{align}
    W_{\alpha\beta}(x,p)=&-\frac{1}{2}\int d^{4}se^{-ip\cdot s}e^{-ie\int_{-1/2}^{1/2}d\lambda s\cdot A(x+\lambda s)}\times\nonumber\\
    &\langle\Omega|\left[\psi_{\alpha}(x+\frac{s}{2}),\bar{\psi}_{\beta}(x-\frac{s}{2})\right]|\Omega\rangle,\label{eq:def_w}
\end{align}
Here, $(x,p)$ are the four-dimensional coordinates and momentum, $A(x)$ is the four-dimensional potential of the background electromagnetic gauge field, $|\Omega\rangle$ is the ground state of the fermion, $\psi(x)$ is the Dirac field operator with $\alpha$ and $\beta$ as the Dirac indices. In this expression, all operators are in the Heisenberg picture.

To study the problem in which we are interested, it is more convenient to deal with a Wigner function that depends on the three-dimensional spatial coordinates, momentum, and time $w(\bm{x},\bm{p},t)$. This is called the equal-time approach, and can be derived by integrating out the zeroth component of $p$ in $W(x,p)$. 

Since both $w(\bm{x},\bm{p},t)$ and $W(x,p)$ are Dirac bispinors, it is straightforward to decompose it as
\begin{align}
    w(\bm{x},\bm{p},t)&=\frac{1}{4}\left[s(\bm{x},\bm{p},t)+i\gamma_{5}s_p(\bm{x},\bm{p},t)+\gamma^{\mu}v_{\mu}(\bm{x},\bm{p},t)+\right.\nonumber\\
    &\left.\gamma^{\mu}\gamma_{5}a_{\mu}(\bm{x},\bm{p},t)+\sigma^{\mu\nu}t_{\mu\nu}(\bm{x},\bm{p},t)\right],\label{eq:decomp}
\end{align}
with $\gamma^\mu$ as the Dirac Matrices, $\gamma_5=i\gamma^0\gamma^1\gamma^2\gamma^3$, $\sigma^{\mu\nu} = (i/2) [\gamma^\mu,\gamma^\nu]$, and $\mu,\nu=0,1,2,3$.

Based on this decomposition, in the remaining part of the paper, we will discuss $w(\bm{x},\bm{p},t)$ in the following representation:
\begin{equation}
    w = \left(s,s_p,v_{0},a_{0},\bm{v},\bm{a},\bm{t}_{1},\bm{t}_{2}\right)^T.\label{eq:rep}
\end{equation}
where $v_0$ and $\bm{v}$ are the temporal and spatial components of $v_{\mu}$; $a_0$ and $\bm{a}$ is the temporal and spatial components of $a_{\mu}$; since $t_{\mu\nu}$ is antisymmetric, $\bm{t}_1$ and $\bm{t}_2$ is defined such that $(\bm{t}_1)_i=2t_{0i}$, $(\bm{t}_2)_i = \epsilon_{ijk} t_{jk}$, $i,j,k = 1,2,3$, and $\epsilon_{ijk}$ as the Levi-Civita symbol. In the below part, we would frequently refer $s,s_p,v_{0},a_{0},\bm{v},\bm{a},\bm{t}_{1},\bm{t}_{2}$ as the components of $w$.

After some algebra with the Dirac equation (see \cite{Kohlfurst2015} for detail), we can obtain the equation of motion under this representation:
\begin{equation}
    \left\{ \begin{array}{l}
        D_{t}s-2\bm{P}\cdot\bm{t}_{1}=0\\
        D_{t}s_p+2\bm{P}\cdot\bm{t}_{2}=-2ma_{0}\\
        D_{t}v_{0}+D_{\bm{x}}\cdot\bm{v}=0\\
        D_{t}a_{0}+D_{\bm{x}}\cdot\bm{a}=2ms_p\\
        D_{t}\bm{v}+D_{\bm{x}}v_{0}+2\bm{P}\times\bm{a}=-2m\bm{t}_{1}\\
        D_{t}\bm{a}+D_{\bm{x}}a_{0}+2\bm{P}\times\bm{v}=0\\
        D_{t}\bm{t}_{1}+D_{\bm{x}}\times\bm{t}_{2}+2\bm{P}s=2m\bm{v}\\
        D_{t}\bm{t}_{2}-D_{\bm{x}}\times\bm{t}_{1}-2\bm{P}s_p=0
    \end{array}\right.,\label{eq:EoMOrdinary}
\end{equation}
where the operators $D_{t}$, $D_{\bm{x}}$, $\bm{P}$ are defined as
\begin{align}
    D_{t}&f(\bm{x},\bm{p},t) = \partial_{t}f(\bm{x},\bm{p},t)\nonumber\\
    &+e\int_{-1/2}^{1/2}d\lambda\bm{E}(\bm{x}+i\lambda\nabla_{\bm{p}},t)\cdot\nabla_{\bm{p}}f(\bm{x},\bm{p},t),\label{eq:Dt}\\
    D_{\bm{x}}&f(\bm{x},\bm{p},t) = \nabla_{\bm{x}}f(\bm{x},\bm{p},t)\nonumber\\
    &+e\int_{-1/2}^{1/2}d\lambda\bm{B}(\bm{x}+i\lambda\nabla_{\bm{p}},t)\times\nabla_{\bm{p}}f(\bm{x},\bm{p},t),\label{eq:Dp}\\
    \bm{P}&f(\bm{x},\bm{p},t) = \bm{p}f(\bm{x},\bm{p},t)\nonumber\\
    &-ie\int_{-1/2}^{1/2}d\lambda\lambda\bm{B}(\bm{x}+i\lambda\nabla_{\bm{p}},t)\times\nabla_{\bm{p}}f(\bm{x},\bm{p},t).\label{eq:P}
\end{align}
Here, $m$ is the mass of the produced fermion, $f \in \{ s,s_p,v_{0},a_{0},\bm{v},\bm{a},\bm{t}_{1},\bm{t}_{2} \}$, $\bm{E}$, $\bm{B}$ are the electric and magnetic fields of $A(x)$, $\nabla_{\bm{x}} = \partial/\partial\bm{x}$, $\nabla_{\bm{p}} = \partial/\partial\bm{p}$.

To solve the equation of motion, boundary condition is required. If we assume that there is no field in the system before $t=0$, then at $t=0$ we can apply the vacuum boundary condition \cite{Berenyi2018}:
\begin{align}
    s(\bm{x},\bm{p},0)=-\frac{2m}{\sqrt{m^{2}+\bm{p}^{2}}},\\
    \bm{v}(\bm{x},\bm{p},0)=-\frac{2\bm{p}}{\sqrt{m^{2}+\bm{p}^{2}}},
\end{align}
while other components of $w(\bm{x},\bm{p},0)$ vanishes.

Finally, once $w(\bm{x},\bm{p},t)$ is solved, we can extract the expectation values of the fermion-related observable. Suppose we have an observable $O$ defined as
\begin{equation}
    O(\bm{x},t)=\frac{1}{2}O^{ab}[\bar{\psi}^{a}(\bm{x},t),\psi^{b}(\bm{x},t)],
\end{equation}
with $O^{ab}$ as some Dirac matrices, then, using the definition of the Wigner function, Eq.~(\ref{eq:def_w}), we have
\begin{equation}
    \langle O(\bm{x},t)\rangle=\int\frac{d^{3}\bm{p}}{(2\pi)^{3}}\mathrm{tr}\left[O^{ab}w^{bc}(\bm{x},\bm{p},t)-O^{ab}w^{bc}(\bm{x},\bm{p},0)\right].
\end{equation}
Here, the trace is taken over the Dirac indices, and the $w(\bm{x},\bm{p},0)$ term is introduced to subtract the vacuum contribution.

With this result, if we choose $O^{ab}=\left(\gamma^\mu\right)^{ab}$, we would find the expectation value of number current density of the produced pairs:
\begin{equation}
    \langle J^{\mu}(t)\rangle=\int\frac{d^{3}\bm{p}}{(2\pi)^{3}}\int d^{3}\bm{x}[v^{\mu}(\bm{x},\bm{p},t)-v^{\mu}(\bm{x},\bm{p},0)],\label{eq:j}
\end{equation}
if we choose $O^{ab}=\left(\gamma^\mu\gamma_5\right)^{ab}$, then we would find the expectation value of the axial current density:
\begin{equation}
    \langle J_{5}^{\mu}(t)\rangle=-\int\frac{d^{3}\bm{p}}{(2\pi)^{3}}\int d^{3}\bm{x} a^{\mu}(\bm{x},\bm{p},t)\label{eq:j5},
\end{equation}
if we choose $O^{ab}=\left( i\gamma_5 \right)^{ab}$, we would find the expectation value of the pseudoscalar condensate \cite{Copinger2018}:
\begin{align}
    \langle \bar{\psi} i\gamma_{5} \psi \rangle (t) &= \int d^3\bm{x} \langle\bar{\psi}(\bm{x},t)i\gamma^{5}\psi(\bm{x},t)\rangle\nonumber\\
    &=- \int\frac{d^{3}\bm{p}}{(2\pi)^{3}}  \int d^3\bm{x} s_p(\bm{x},\bm{p},t).\label{eq:pseu-cond}
\end{align}
Finally, through a more complicated discussion (see \cite{Hebenstreit2011-2} for detail), we have the number of the fermions:
\begin{align}
    \langle N(t) \rangle=&\int\frac{d^{3}\bm{p}}{(2\pi)^{3}}\int d^{3}\bm{x} \frac{1}{\epsilon_p}
    \left[ms(\bm{x},\bm{p},t)+\bm{p}\cdot\bm{v}(\bm{x},\bm{p},t) \right.\nonumber\\
    &\left.-ms(\bm{x},\bm{p},0)-\bm{p}\cdot\bm{v}(\bm{x},\bm{p},0)\right],\label{eq:partical-number}
\end{align}
with $\epsilon_p=(\bm{p}^2+m^2)^{1/2}$.

\section{General Solution of the Equation of Motion\protect\label{sec:gen-sol}}

In the equation of motion, Eq.~(\ref{eq:EoMOrdinary}), the operators $D_{t}$, $D_{\bm{x}}$, and $\bm{P}$ [Eqs.~(\ref{eq:Dt})-(\ref{eq:P})] has $\nabla_{\bm{p}}$ operator in the spatial coordinates of $\bm{E}$ and $\bm{B}$, which makes solving the equations rather complex. To simplify these operators, we perform the following Fourier transformation $\tilde{f}(\bm{q},\bm{p},t)=\int d^{3}\bm{y}e^{-i\bm{p}\cdot\bm{y}}\left[\int d^{3}\bm{x}e^{-i\bm{q}\cdot\bm{x}}\int d^{3}\bm{p}/(2\pi)^{3} e^{i\bm{p}\cdot\bm{y}}f(\bm{x},\bm{p},t)\right]$ on each side of the equation of motion. The result is a matrix equation for the  Fourier transformed Wigner function $\tilde{w}(\bm{q},\bm{p},t)$:
\begin{equation}
    \partial_{t}\tilde{w}(\bm{q},\bm{p},t)+A(\bm{q},\bm{p})\tilde{w}(\bm{q},\bm{p},t)=S(t)\tilde{w}(\bm{q},\bm{p},t),\label{eq:EoMGen}
\end{equation}
with $A(\bm{q},\bm{p})$ defined as
\begin{equation}
    A(\bm{q},\bm{p})\equiv\left(\begin{array}{cccccccc}
        0 & 0 & 0 & 0 & 0 & 0 & -2\bm{p}\cdot & 0\\
        0 & 0 & 0 & 2m & 0 & 0 & 0 & 2\bm{p}\cdot\\
        0 & 0 & 0 & 0 & i\bm{q}\cdot & 0 & 0 & 0\\
        0 & -2m & 0 & 0 & 0 & i\bm{q}\cdot & 0 & 0\\
        0 & 0 & i\bm{q} & 0 & 0 & 2\bm{p}\times & 2m & 0\\
        0 & 0 & 0 & i\bm{q} & 2\bm{p}\times & 0 & 0 & 0\\
        2\bm{p} & 0 & 0 & 0 & -2m & 0 & 0 & i\bm{q}\times\\
        0 & -2\bm{p} & 0 & 0 & 0 & 0 & -i\bm{q}\times & 0
        \end{array}\right)\label{eq:A},
\end{equation}
and $S(t)$ defined as
\begin{equation}
    \resizebox{0.9\linewidth}{!}{
    $
        \hat{S}(t)\equiv\left(\begin{array}{cccccccc}
            -S_{t} & 0 & 0 & 0 & 0 & 0 & 2S_{\bm{p}}\cdot & 0\\
            0 & -S_{t} & 0 & 0 & 0 & 0 & 0 & -2S_{\bm{p}}\cdot\\
            0 & 0 & -S_{t} & 0 & -S_{\bm{x}}\cdot & 0 & 0 & 0\\
            0 & 0 & 0 & -S_{t} & 0 & -S_{\bm{x}}\cdot & 0 & 0\\
            0 & 0 & -S_{\bm{x}} & 0 & -S_{t} & -2S_{\bm{p}}\times & 0 & 0\\
            0 & 0 & 0 & -S_{\bm{x}} & -2\hat{S}_{\bm{p}}\times & -S_{t} & 0 & 0\\
            -2S_{\bm{p}} & 0 & 0 & 0 & 0 & 0 & -S_{t} & -S_{\bm{x}}\times\\
            0 & 2S_{\bm{p}} & 0 & 0 & 0 & 0 & S_{\bm{x}}\times & -S_{t}
            \end{array}\right),
    $
    }\label{eq:S}
\end{equation}
\begin{align}
    S_{t}\tilde{f}(\bm{q},\bm{p},t)=&e\int\frac{d^{3}\bm{s}}{(2\pi)^{3}}\int_{-1/2}^{1/2}d\lambda\times\nonumber\\
    &\tilde{\bm{E}}(\bm{s},t)\cdot\nabla_{\bm{p}}\tilde{f}(\bm{q}-\bm{s},\bm{p}-\lambda\bm{s},t),\label{eq:St}\\
    S_{\bm{x}}\tilde{f}(\bm{q},\bm{p},t)=&e\int\frac{d^{3}\bm{s}}{(2\pi)^{3}}\int_{-1/2}^{1/2}d\lambda\times\nonumber\\
    &\tilde{\bm{B}}(\bm{s},t)\times\nabla_{\bm{p}}\tilde{f}(\bm{q}-\bm{s},\bm{p}-\lambda\bm{s},t),\label{eq:Sx}\\
    S_{\bm{p}}\tilde{f}(\bm{q},\bm{p},t)=&-ie\int\frac{d^{3}\bm{s}}{(2\pi)^{3}}\int_{-1/2}^{1/2}d\lambda\lambda\times\nonumber\\
    &\tilde{\bm{B}}(\bm{s},t)\times\nabla_{\bm{p}}\tilde{f}(\bm{q}-\bm{s},\bm{p}-\lambda\bm{s},t),\label{eq:Sp}
\end{align}
where $\tilde{f}(\bm{q},\bm{p},t)$ is some components of $\tilde{w}(\bm{q},\bm{p},t)$; meanwhile, $\tilde{\bm{E}}(\bm{s},t)=\int d^3\bm{x} \bm{E}(\bm{x},t) e^{-i\bm{s}\cdot\bm{x}}$, $\tilde{\bm{B}}(\bm{s},t)=\int d^3\bm{x} \bm{B}(\bm{x},t) e^{-i\bm{s}\cdot\bm{x}}$.

The initial condition of this Fourier transformed equation of motion becomes
\begin{align}
    \tilde{s}(\bm{q},\bm{p},0)&=\frac{-2m}{\sqrt{m^{2}+\bm{p}^{2}}}(2\pi)^{3}\delta(\bm{q})\equiv s(\bm{p},0)(2\pi)^3\delta(\bm{q}),\label{eq:ini1}\\
    \tilde{\bm{v}}(\bm{q},\bm{p},0)&=\frac{-2\bm{p}}{\sqrt{m^{2}+\bm{p}^{2}}}(2\pi)^{3}\delta(\bm{q})\equiv \bm{v}(\bm{p},0)(2\pi)^3\delta(\bm{q})\label{eq:ini2}.
\end{align}

Up to this step, Eq.~(\ref{eq:EoMGen}) is completely general for any field configuration. Now, we apply the homogeneous field setup Eqs.~(\ref{eq:bkgr_E}),(\ref{eq:bkgr_B}) such that $w(\bm{x},\bm{p},t)$ becomes $\bm{x}$ independent. In this situation, we can define the spatially averaged Wigner function $w(\bm{p},t) = (1/V) \int d^{3}\bm{x}w(\bm{x},\bm{p},t)$ (where $V$ is the volume of the system) and proof the equation of motion of $w(\bm{p},t)$ is
\begin{equation}
    \partial_{t}w(\bm{p},t)+A(\bm{p})w(\bm{p},t)=S(t)w(\bm{p},t),\label{eq:EoM_uniform}
\end{equation}
\begin{equation}
    A(\bm{p})w(\bm{p},t)=\left(\begin{array}{c}
        -2\bm{p}\cdot\bm{t}_{1}(\bm{p},t)\\
        2ma_{0}(\bm{p},t)+2\bm{p}\cdot\bm{t}_{2}(\bm{p},t)\\
        0\\
        -2ms_p(\bm{p},t)\\
        2\bm{p}\times\bm{a}(\bm{p},t)+2m\bm{t}_{1}(\bm{p},t)\\
        2\bm{p}\times\bm{v}(\bm{p},t)\\
        2\bm{p}s(\bm{p},t)-2m\bm{v}(\bm{p},t)\\
        -2\bm{p}s_p(\bm{p},t)
    \end{array}\right),
\end{equation}
\begin{align}
    S(t)w(\bm{p},t)&=-eE(t)\partial_{p_{z}}w(\bm{p},t) -\nonumber\\
        &-eB(t)\left(\begin{array}{c}
            0\\
            0\\
            (\bm{e}_{y}\partial_{p_{x}}-\bm{e}_{x}\partial_{p_{y}})\cdot\bm{v}(\bm{p},t)\\
            (\bm{e}_{y}\partial_{p_{x}}-\bm{e}_{x}\partial_{p_{y}})\cdot\bm{a}(\bm{p},t)\\
            (\bm{e}_{y}\partial_{p_{x}}-\bm{e}_{x}\partial_{p_{y}})v_{0}(\bm{p},t)\\
            (\bm{e}_{y}\partial_{p_{x}}-\bm{e}_{x}\partial_{p_{y}})a_{0}(\bm{p},t)\\
            (\bm{e}_{y}\partial_{p_{x}}-\bm{e}_{x}\partial_{p_{y}})\times\bm{t}_{2}(\bm{p},t)\\
            -(\bm{e}_{y}\partial_{p_{x}}-\bm{e}_{x}\partial_{p_{y}})\times\bm{t}_{1}(\bm{p},t)
        \end{array}\right).\label{eq:S_uniform}
\end{align}
Here, $\bm{e}_x$, $\bm{e}_y$ are the unit vector in $x$ and $y$ direction.

From the above equation, it is straightforward to see that $[S(t_1),S(t_2)]=0$, so with $\mathcal{T}[\cdot]$ as the time-order product, we can write down $\mathcal{T}[S(t_1)S(t_2)]=S(t_1)S(t_2)$. Then, the formal solution of the equation of motion becomes:
\begin{equation}
    w(\bm{p},t) = e^{ - \left[t A(\bm{p}) + \int_{0}^t d\bar{t} S(\bar{t}) \right]} w(\bm{p},0),\label{eq:general_solution}
\end{equation}
with $ w(\bm{p},0) = \left(s(\bm{p},0),0,0,0,\bm{v}(\bm{p},0),0,0,0\right)^T$ as those given in Eqs.~(\ref{eq:ini1}),(\ref{eq:ini2}).

Furthermore, based on the definition of $w(\bm{p},t)$ and Eqs.~(\ref{eq:j})-(\ref{eq:partical-number}), it is straightforward to define the charge current distribution, the axial current distribution, the pseudoscalar condensate distribution, and the particle number distribution on the momentum space, respectively, which are
\begin{align}
    j^\mu(\bm{p},t) &= v^{\mu}(\bm{p},t) - v^{\mu}(\bm{p},0),\label{eq:def-j}\\
    j_5^\mu(\bm{p},t) &= -a^\mu(\bm{p},t),\label{eq:def-j5}\\
    \langle \bar{\psi} i\gamma_{5} \psi \rangle (\bm{p},t) &= -s_p(\bm{p},t),\label{eq:def-p}\\
    n(\bm{p},t) &= \frac{m(s\left(\bm{p},t) - s(\bm{p},0)\right) + \bm{p} \cdot \left(\bm{v}(\bm{p},t) - \bm{v}(\bm{p},0)\right)}{\epsilon_p}\label{eq:def-n}.
\end{align}
With these definitions, the expectation values of the corresponding observable can be obtained from the following integral:
\begin{equation}
    \langle X(t) \rangle = V \int \frac{d^3\bm{p}}{(2\pi)^3} x(\bm{p},t),\label{eq:spatial-averages}
\end{equation}
with $x=j,j_5,\langle \bar{\psi} i\gamma_{5} \psi \rangle, n$, $X=J,J_5,\langle \bar{\psi} i\gamma_{5} \psi \rangle ,N$, respectively.

\section{The Excitation Stage\protect\label{sec:ex-stage}}

In this section, we discuss the time span $0<t<\tau$, when the external fields have not vanished, and refer to it as the excitation stage. In the excitation stage, since $\tau$ is small, we can expand the function $\exp(\cdots)$ in Eq.~(\ref{eq:general_solution}) with respect to different orders of $t$. The result shows that, at $t=\tau$, there is
\begin{align}
    s(\bm{p},\tau) - s(\bm{p},0)&=\frac{2m}{\epsilon_{p}^{3}}eE_{0}\tau p_{z}+\nonumber\\
    &+\frac{m}{\epsilon_{p}^{5}}e^{2}E_{0}^{2}\tau^{2}(\epsilon_{p}^{2}-3p_{z}^{2})+O(\tau^{3}),\label{eq:s-excitation}
\end{align}
\begin{align}
    \bm{v}(\bm{p},\tau) - \bm{v}(\bm{p},0) &= \frac{2}{\epsilon_{p}^{3}}eE_{0}\tau\left(\begin{array}{c}
        p_{x}p_{z}\\
        p_{y}p_{z}\\
        p_{z}^{2}-\epsilon_{p}^{2}
        \end{array}\right)+\nonumber\\
    &+\frac{1}{\epsilon_{p}^{5}}e^{2}E_{0}^{2}\tau^{2}\left(\begin{array}{c}
        p_{x}(\epsilon_{p}^{2}-3p_{z}^{2})\\
        p_{y}(\epsilon_{p}^{2}-3p_{z}^{2})\\
        3p_{z}(\epsilon_{p}^{2}-p_{z}^{2})
        \end{array}\right)
    +O(\tau^{3})\label{eq:v-excitation}.
\end{align}
Then, according to Eq.~(\ref{eq:def-n}), the particle number distribution becomes
\begin{align}
    n(\bm{p},\tau) = e^{2}E_{0}^{2}\tau^{2}\frac{1}{\epsilon_{p}^{2}}(1-\frac{p_{z}^{2}}{\epsilon_{p}^{2}})+O(\tau^{3}).\label{eq:energy}
\end{align}

Similarly, we can derive the axial charge distribution $\rho_5(\bm{p},t) = j_5^0(\bm{p},t)$ that is defined by Eq.~(\ref{eq:def-j5}), which is
\begin{align}
    \rho_5(\bm{p},\tau) =\frac{2}{3}\frac{1}{\epsilon_{p}^{3}}e^{2}E_{0}B_{0}\tau^{3}(\epsilon_{p}^{2}+p_{z}^{2}+m^{2})+O(\tau^{4}).\label{eq:a0_excitation}
\end{align}
Therefore the electromagnetic pulses satisfying $\bm{E}\parallel\bm{B}$ indeed produces chirality as expected.

On the other hand, the pseudoscalar condensate distribution $\langle \bar{\psi} i\gamma_{5} \psi \rangle (\bm{p},t)$ defined by Eq.~(\ref{eq:def-p}) is:
\begin{equation}
    \langle \bar{\psi} i\gamma_{5} \psi \rangle (\bm{p},\tau) = -\frac{2m}{3\epsilon_{p}}e^{2}E_0B_0\tau^{4} + O(\tau^5).
\end{equation}
Hence, $\langle \bar{\psi} i\gamma_{5} \psi \rangle(\bm{p},\tau)\sim O(\tau^4)$, while $\rho_5(\bm{p},\tau)\sim O(\tau^3)$. For short electromagnetic pulses, the excited chiral charge is significantly larger than the pseudoscalar condensate. As we would see in Sec. \ref{sec:result}, this property leads to very interesting outcomes.

For other components of $w(\bm{p},\tau)$, the results up to the fourth order of $\tau$ are listed in Appendix \ref{sec:to4}.

\section{The Evolution Stage\protect\label{sec:free-stage}}

In this section, we discuss the time span $t>\tau$, when the external fields vanish and the system evolves freely, and refer to it as the evolution stage. At the evolution stage, the equation of motion, Eq.(\ref{eq:EoM_uniform}), decouples into three groups of independent equations, which are as follows:
\begin{enumerate}
    \item The charge conservation equation:
    \begin{equation}
        \partial_t v_0 = 0.
    \end{equation}
    From Eq.~(\ref{eq:def-j}), we can immediately show that this equation guarantees that the electric charge distribution of the fermions $\rho(\bm{p},t)$ does not change during the evolution stage. Combining this fact with $v_0^{(n)}(\bm{p},\tau)=0$, $n=1,2,3,4$ (see Appendix \ref{sec:to4}), we would know that no net electric charge is produced in multiphoton pair production, which is expected since the fermions are produced as particle-antiparticle pairs that have opposite charge.
    \item The particle number evolution equations:
    \begin{equation}
       \left\{ \begin{array}{l}
            \partial_{t}s-2\bm{p}\cdot\bm{t}_{1}=0\\
            \partial_{t}\bm{v}+2\bm{p}\times\bm{a}+2m\bm{t}_{1}=0\\
            \partial_{t}\bm{a}+2\bm{p}\times\bm{v}=0\\
            \partial_{t}\bm{t}_{1}+2\bm{p}s-2m\bm{v}=0
            \end{array}\right.\label{eq:eqs_num}.
    \end{equation}
    These equations give the particle number distribution $n(\bm{p},t)$.
    \item The chirality evolution equations:
    \begin{equation}
        \left\{ \begin{array}{l}
            \partial_{t}s_p+2ma_{0}+2\bm{p}\cdot\bm{t}_{2}=0\\
            \partial_{t}a_{0}-2ms_p=0\\
            \partial_{t}\bm{t}_{2}-2\bm{p}s_p=0
            \end{array}\right.\label{eq:eqs_chiral}.
    \end{equation}
    These equations give the chiral charge distribution $\rho_5(\bm{p},t)$ and the pseudoscalar condensate distribution $\langle \bar{\psi} i\gamma_{5} \psi \rangle(\bm{p},t)$.
\end{enumerate}

First, let us solve Eq.~(\ref{eq:eqs_num}) using $s(p,\tau)$, $\bm{v}(p,\tau)$, $\bm{a}(\bm{p},\tau)$, $\bm{t}_1(\bm{p},\tau)$ as initial conditions. After some simplification based on Laplace transformation, we found out that
\begin{equation}
    n(\bm{p},t) = n(\bm{p},\tau) = e^{2}E_{0}^{2}\tau^{2}\frac{1}{\epsilon_{p}^{2}}(1-\frac{p_{z}^{2}}{\epsilon_{p}^{2}})+O(\tau^{3}),
\end{equation}
so the particle number does not change after the external fields are switched off. Considering that the vacuum becomes stable again and no new pairs can be produced afterwards, and that we ignore the interaction among the produced pairs, this conclusion is completely natural. Therefore, later we will denote $n(\bm{p})\equiv n(\bm{p},t)$.

Next, we switch to the chirality evolution equations, Eq.~(\ref{eq:eqs_chiral}), that will lead much more nontrivial results. We use the initial condition $a_0(\bm{p},\tau)$, $s_p(\bm{p},\tau)$, $\bm{t}_2(\bm{p},\tau)$, and solve the equations by the Laplace transformation. The resulting $s_p(\bm{p},t)$, $a_0(\bm{p},t)$ is
\begin{align}
    s_{p}(\bm{p},t)&=s_p(\bm{p},\tau)\cos(2\epsilon_{p}t)-\nonumber\\
    &-\frac{1}{\epsilon_{p}}\left(ma_{0}(\bm{p},\tau)+\bm{p}\cdot\bm{t}_{2}(\bm{p},\tau)\right)\sin(2\epsilon_{p}t),\label{eq:pseu-cond-oscillation}
\end{align}
\begin{align}
    a_{0}(\bm{p},t)&=a_{0}(\bm{p},\tau)+\frac{m}{\epsilon_{p}}s_p(\bm{p},\tau)\sin(2\epsilon_{p}t)+\nonumber\\
    &+\frac{m}{\epsilon_{p}^{2}}\left(ma_{0}(\bm{p},\tau)+\bm{p}\cdot\bm{t}_{2}(\bm{p},\tau)\right)\left(\cos(2\epsilon_{p}t)-1\right).
\end{align}
Therefore, in the evolution stage, $s_p(\bm{p},t)$, $a_0(\bm{p},t)$ experience oscillation of frequency $2\epsilon_p$ and do not reach the maximum value at the same time. This oscillating behavior occurs because in Eq.~(\ref{eq:eqs_chiral}),  $a_0$, $s_p$, and $\bm{t}_2$ are coupled with each other by their first-order time derivatives.

In the first glimpse, this oscillation may seem strange, as the intuitive expectation is that after the external fields vanish, the chiral charge will undergo a monotonic decay with respect to time if the particles are massive. However, the existence of this oscillating behavior can be predicted even without the DHW formalism, since $\partial_{t}a_{0}-2ms_p=0$ is the obvious consequence of the axial Ward identity $\partial_\mu j_5^\mu = e^2/(2\pi^2) \bm{E}\cdot\bm{B} + 2m \bar{\psi} i\gamma_5 \psi$, and $\partial_{t}s_p+2ma_{0}+2\bm{p}\cdot\bm{t}_{2}=0$ can be derived directly from the free-particle Dirac equation as what we have done in Appendix \ref{sec:dev-chiral}.

Now, let us compute the time average of $s_p(\bm{p},t)$ and $a_0(\bm{p},t)$, denoted as $s_p(\bm{p})$ and $a_0(\bm{p})$, respectively. We find $s_p(\bm{p})=0$ and
\begin{equation}
    a_{0}(\bm{p})=a_{0}(\bm{p},\tau)-\frac{m}{\epsilon_{p}^{2}}\left(ma_{0}(\bm{p},\tau)+\bm{p}\cdot\bm{t}_{2}(\bm{p},\tau)\right).\label{eq:a0-avg}
\end{equation}
This result shows that when $m\to 0$, $a_0(\bm{p})\to a_0(\bm{p},\tau)$; when $m\to\infty$, $a_0\to 0$. Thus, during the evolution stage, the chiral charge is suppressed by the mass fermion, as expected.

Finally, we substitute the explicit expression of $a_0(\bm{p},\tau)$ and $\bm{t}_2(\bm{p},\tau)$ in Appendix \ref{sec:to4} in Eq.~(\ref{eq:a0-avg}), and find out that the time-averaged value of $\rho_5(\bm{p},t)$ to be
\begin{equation}
    \rho_5(\bm{p})=\frac{2}{3}\frac{1}{\epsilon_{p}^{3}}e^{2}E_{0}B_{0}\tau^{3}(\epsilon_{p}^{2}+p_{z}^{2}-m^{2})+O(\tau^{4})\label{eq:n5-final}.
\end{equation}
As we will see in the next section, this result will lead to an interesting spectrum structure of $\rho_5(\bm{p})$.

\section{Results and Discussion\protect\label{sec:result}}

\begin{figure*}
    \begin{subfigure}{0.49\linewidth}
        \includegraphics[width=\linewidth]{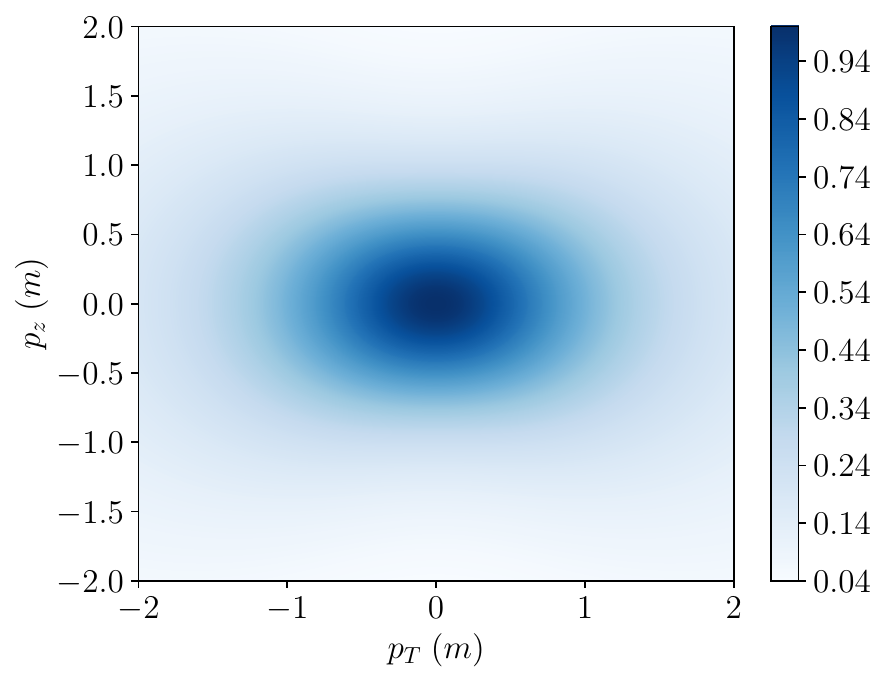}
        \caption{\label{fig:n-plots-contour}}
    \end{subfigure}
    \begin{subfigure}{0.49\linewidth}
        \includegraphics[width=\linewidth]{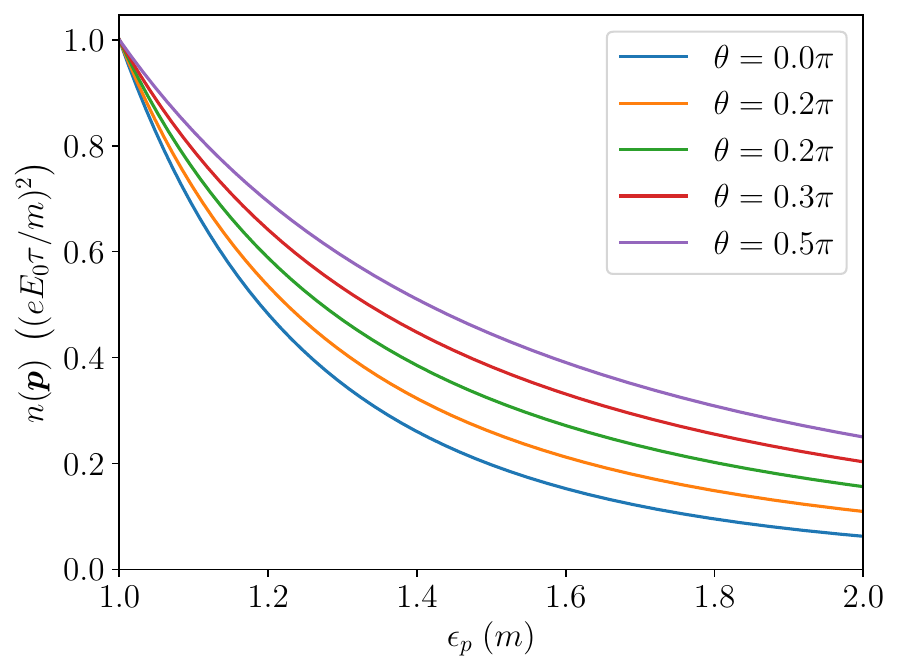}
        \caption{\label{fig:n-plots-epsilonp}}
    \end{subfigure}
    \caption{Particle number distribution $n(\bm{p})$. (a) The contour plots of $n(\bm{p})$ with respect to $p_T=(p_x^2+p_y^2)^{1/2}$ and $p_z$, with the magnitude normalized by $(e E_0 \tau/m)^2$. (b) $n(\bm{p})$ with respect to $\epsilon_p=(\bm{p}^2+m^2)^{1/2}$ in different $\theta = \arctan{(p_T/p_z)}$. The $O(\tau^3)$ part of the results in (a),(b) is neglected. \label{fig:n-plots}}
\end{figure*}

First, we discuss the particle number distribution. Taking into account the rotational invariance of the system in $z$ direction, we make the contour plot of $n(\bm{p})$ with respect to the transverse momentum $p_T=(p_x^2+p_y^2)^{1/2}$ and the longitudinal momentum $p_z$, as shown in Fig.~\ref{fig:n-plots-contour}. Similarly, $n(\bm{p})$ with respect to $\theta = \arctan{(p_T/p_z)}$ and particle energy $\epsilon_p$ are plotted in Fig.~\ref{fig:n-plots-epsilonp}.

From these figures we can see, the amount of pairs drops with the increase of the particle energy $\epsilon_p$. This is quite natural because the external fields that excite the pairs are pulses, which do not favor any particular energy, so the distribution with respect to $\epsilon_p$ exhibits a shape similar to the initial distribution in Eqs.~(\ref{eq:ini1}),(\ref{eq:ini2}). Also, the pairs are emitted anisotropically, with more pairs emitted along the transverse ($xOy$) direction.

\begin{figure*}
    \begin{subfigure}{0.49\linewidth}
        \includegraphics[width=\linewidth]{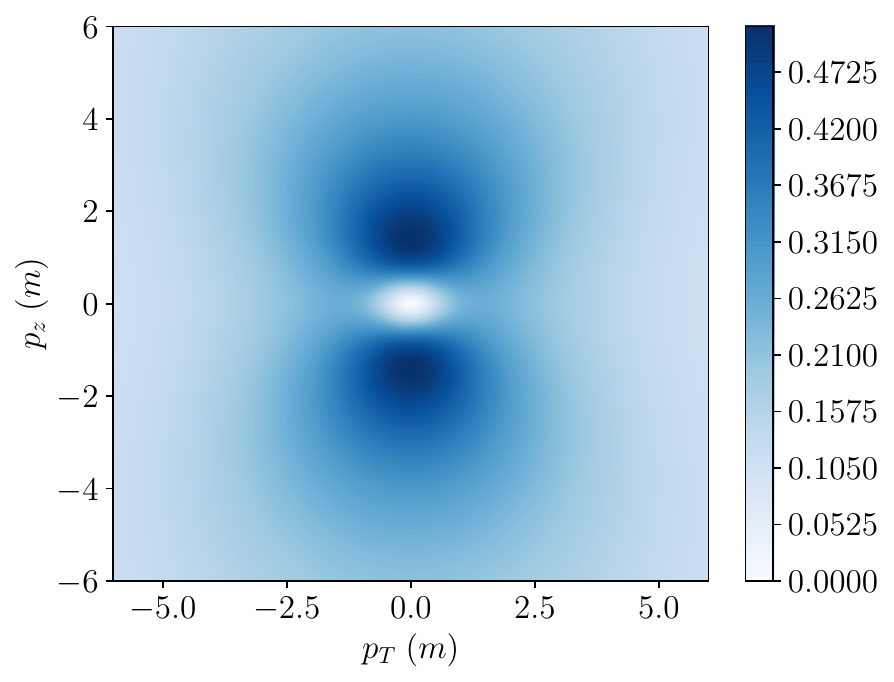}
        \caption{\label{fig:n5-plots-contour}}
    \end{subfigure}
    \begin{subfigure}{0.49\linewidth}
        \includegraphics[width=\linewidth]{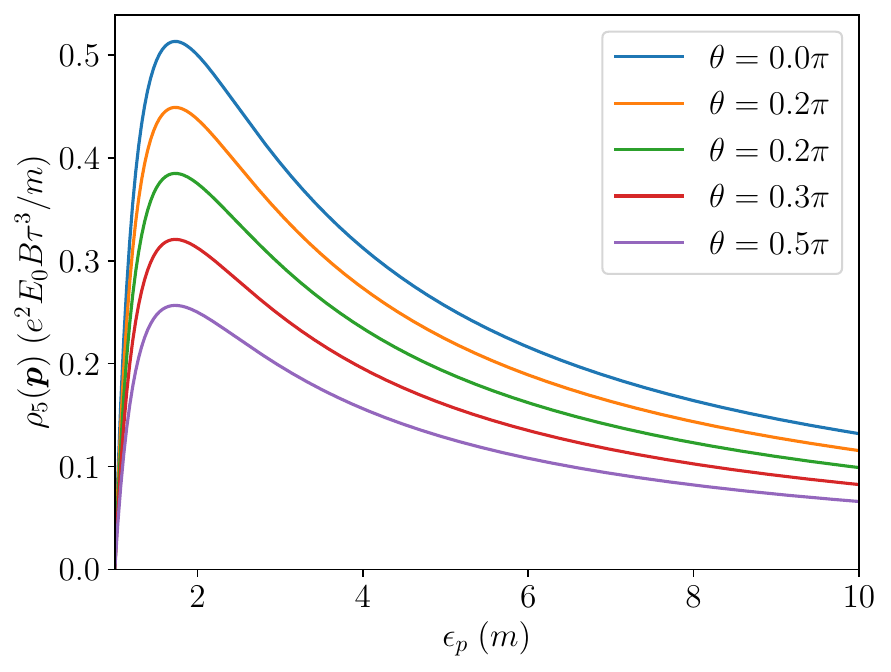}
        \caption{\label{fig:n5-plots-epsilonp}}
    \end{subfigure}
    \caption{Chiral charge distribution $\rho_5(\bm{p})$. (a) The contour plots of $\rho_5(\bm{p})$ with respect to $p_T=(p_x^2+p_y^2)^{1/2}$ and $p_z$, with the magnitude normalized by $e^2 E_0 B_0 \tau^3/m$. (b) $\rho_5(\bm{p})$ with respect to $\epsilon_p=(\bm{p}^2+m^2)^{1/2}$ at different $\theta = \arctan{(p_T/p_z)}$. The $O(\tau^4)$ part of the results in (a),(b) are neglected. \label{fig:n5-plots}}
\end{figure*}

Then, we discuss the distribution of the axial charge $\rho_5(\bm{p})$ given by Eq.~(\ref{eq:n5-final}), see Fig.~\ref{fig:n5-plots-contour} for the contour plot of $\rho_5(\bm{p})$ with respect to $p_T$ and $p_z$ , and Fig.~\ref{fig:n5-plots-epsilonp} for $\rho_5(\bm{p})$ with respect to $\theta$ and $\epsilon_p$.

These figures show two interesting characteristics:

First of all, the distribution of the chiral charge does not drop monotonically with the increase of $\epsilon_p$; on the contrary, there is a nonzero energy value at which the production of chirality is maximized. Furthermore, this energy value does not depends on either the direction of emission $\theta$, or the profile of the external electromagnetic fields. Instead, we can derive from Eq.~(\ref{eq:n5-final}) that up to the third order of $\tau$, the energy value is always $\epsilon_p = \sqrt{3} m$. This relation establishes that the nonmonotonic behavior is an intrinsic property of fermions themselves. In the later part of this section, we will discuss this behavior in detail.

Apart from nonmonotonic behavior discussed above, Fig.~\ref{fig:n5-plots} also tells us that the chirality production is maximized in the $z$ direction. On the other hand, Fig.~\ref{fig:n-plots} shows that at $z$ direction, the number of produced pairs is minimized, so the chiral charge per particle would be quite high. In the future experiments, we may be able to find highly chiral particles in this direction.

\begin{figure}
    \includegraphics[width=\linewidth]{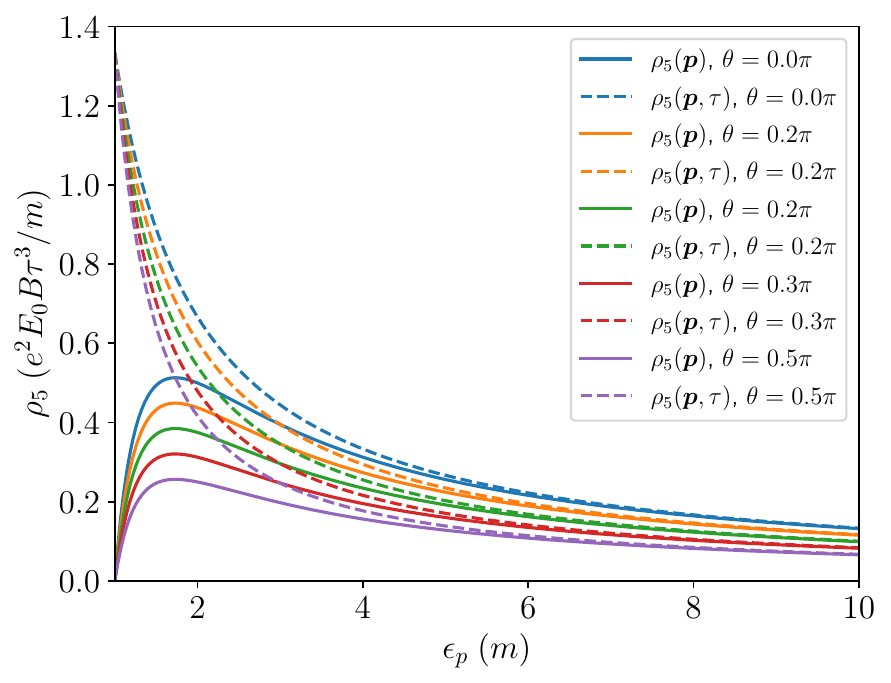}
    \caption{The chiral charge distribution at the end of the excitation stage, $\rho_5(\bm{p},\tau)$, and the time-averaged chiral charge distribution in the evolution stage, $\rho_5(\bm{p})$, at different $\theta=\arctan{[(p_x^2+p_y^2)^{1/2}/p_z]} $ and $\epsilon_p=(\bm{p}^2+m^2)^{1/2}$. The $O(\tau^4)$ part of the results are neglected.}
    \label{fig:compare-n5}
\end{figure}

In the remaining part of this section, let us focus on discussing the mechanism of the nonmonotonic behavior in Fig.~\ref{fig:n5-plots}. For this purpose, we plot the the chiral charge distribution at the end of the excitation stage, $\rho_5(\bm{p},\tau)$, given by Eq.~(\ref{eq:a0_excitation}), as well as the time-averaged chiral charge distribution in the evolution stage $\rho_5(\bm{p})$, given by Eq.~(\ref{eq:n5-final}), in the same figure, Fig.~\ref{fig:compare-n5}.

This figure shows that the nonmonotonic behavior does not occur just after the vanishing of the external electromagnetic field, but emerges during the latter evolution of the system. This remind us of an observable that exhibits similar behavior -- the pseudoscalar condensate $p(\bm{p},t)$. As discussed in Sec. \ref{sec:ex-stage}, at the end of the excitation stage, the pseudoscalar condensate is much smaller than the chiral charge. On the other hand, as calculated in Eq.~(\ref{eq:pseu-cond-oscillation}), in the evolution stage, an oscillating pseudoscalar condensate whose magnitude is comparable to that of the chiral charge density would occur.

The interesting thing is, as shown by the equation $\partial_{t}a_{0}-2ms_p=0$ in Eq.~(\ref{eq:eqs_chiral}), the increase of pseudoscalar condensate is actually made possible by transforming part of the chiral charge into the pseudoscalar condensate. As a result of this transformation, in Fig.~\ref{fig:compare-n5}, at all values of $\epsilon_p$ and $\theta$, the average magnitude of the chiral charge $\rho_5(\bm{p})$ is suppressed comparing with that at the end of the excitation stage, $\rho_5(\bm{p},\tau)$.

In fact, it is this suppression of chiral charge that leads to the nonmonotonic behavior in Fig.~\ref{fig:n5-plots}. The reason is this: according to $\partial_t a_0-2ms_p=0$ and the fact that $a_0$ is oscillating with frequency $2\epsilon_p$, the coupling between $a_0$ and $s_p$ is proportional to $m/\epsilon_p$, so, for low energy particles, a large portion of the chirality will be transformed into peseudoscalar condensate, while for high energy particles, the ratio is much less; on the other hand, since the external fields are pulse shaped, more fermion pairs and hence more chiral charge will be produced at low energy during the excitation stage, as shown in Fig.~\ref{fig:compare-n5}. These two mechanisms would compete with each other. As a result, an optimized energy value for chirality production must be achieved in the intermediate energy regime.

\section{Conclusions and Perspectives\protect\label{sec:concl}}

In this paper, we have studied the production and evolution of the chiral charge in vacuum excited by spatially homogeneous external electromagnetic pulses that satisfies $\bm{E}\parallel\bm{B}$, which is a simplified model for the laser pulses in multiphoton pair production experiments. Based on the DHW formalism, we analytically solves the model to obtain the Wigner function of the fermion pairs excited by the fields, and discovered the following:
\begin{enumerate}
    \item The largest portion of the chiral charge is owned by the fermions propagating in the direction parallel to the electric and magnetic fields, whereas the number of produced fermions minimizes in the same direction.
    \item After the external fields vanish, if the fermions are massive, then part of the chiral charge will be transformed into a rapidly oscillating pseudoscalar condensate\label{item:finding2}.
    \item As a result of the oscillation, the average chiral charge will be suppressed comparing with the amount the fermions have obtained from the external fields before the fields vanish\label{item:finding3}.
    \item Most interestingly, this suppression would lead to a nonmonotonic behavior on the $\epsilon_p$-$\rho_5$ spectrum. An optimized energy value $\epsilon_p = \sqrt{3} m$, which is irrelevant to both the field profile and the direction of particle emission, would allow particles with this energy to own the largest amount of chirality \label{item:finding4}.
\end{enumerate}
To our knowledge, findings~\ref{item:finding2}--\ref{item:finding4} have never been discussed in literature.

For findings~\ref{item:finding2}--\ref{item:finding4}, our intuitive physics picture is this: the chiral charge and pseudoscalar condensate are mutually coupled to the first-order time derivative of each other, just like the kinetic and potential energy of a harmonic oscillator, hence the oscillation occurs. At the same time, the oscillation starts at a initial state where chiral charge is much larger than pseudoscalar condensate, so after taking the time-average, the chiral charge is suppressed. Furthermore, the particles with large $m/\epsilon_p$ are more likely to lose chirality; however, pulsed $\mathbf{E}$ and $\mathbf{B}$ also excites more low-energy particles than the high-energy ones. These two tendencies compete with each other, thus nonmonotonic behavior occurs in the intermediate energy regime.

These findings could lead to interesting applications in the future multiphoton pair production experiments that involve the production of chirality. For example, in the experiments, different types of particles might be produced in the same event; but since we know for each type of product, a peak $\epsilon_p=\sqrt{3}m$ would occur on the chirality spectrum, we can use the peaks to identify different types of products with different masses, even before separating the products with experimental measures. This would allow us to extract more information about the multiphoton pair production process.

Because of its usefulness, in the future, we plan to extend our model to include spatially inhomogeneous $\bm{E}(\bm{x},t)$ and $\bm{B}(\bm{x},t)$. At this situation, one possible new phenomenon is that the oscillation of pseudoscalar condensate and chiral charge would become a wave. If this is true, then the wave may produce interesting outcomes in the future photon pair production experiment, which is worth further investigation. Also, the relation between this wave and the chiral magnetic wave, which is under wide discussion in the context of high energy heavy ion collision and neutron star physics \cite{Kharzeev2011,Adam2016,Hanai2022}, is another direction that could lead to fruitful outcomes. 

\begin{acknowledgments}
The author would like to thank Professor Kenji Fukushima of the University of Tokyo for valuable discussions on the results of this paper.
\end{acknowledgments}

\appendix
\section{$w(\mathbf{p},\tau)$ up to the 4th Order of $\tau$ \label{sec:to4}}

After conducting expansion over $\tau$, Eq.~(\ref{eq:general_solution}) at $t=\tau$ can be written as
\begin{equation}
    w(\bm{p},\tau) = \sum_{n=0}^{\infty} w^{(n)}(\bm{p},\tau),
\end{equation}
with $w^{(n)}(\bm{p},\tau)\propto\tau^n$.

The zeroth order result $w^{(0)}(\bm{p},\tau)$ is the initial condition, whose nonzero components are $s(\bm{p},0)$ and $\bm{v}(\bm{p},0)$, given in Eqs.~(\ref{eq:ini1}),(\ref{eq:ini2}).

The first order result is
\begin{align}
    s^{(1)}(\bm{p},\tau)&=\frac{2m}{\epsilon_{p}^{3}}eE_{0}\tau p_{z},\label{eq:s-order1}\\
    \bm{v}^{(1)}(\bm{p},\tau)&=\frac{2}{\epsilon_{p}^{3}}eE_{0}\tau\left(\begin{array}{c}
        p_{x}p_{z}\\
        p_{y}p_{z}\\
        p_{z}^{2}-\epsilon_{p}^{2}
        \end{array}\right)\label{eq:v-order1}.
\end{align}
Other components are zeros.

The second order result is
\begin{align}
    s^{(2)}(\bm{p},\tau)&=\frac{m}{\epsilon_{p}^{5}}e^2E_0^2\tau^2(\epsilon_{p}^{2}-3p_{z}^{2}),\label{eq:s_order2}\\
    \bm{v}^{(2)}(\bm{p},\tau)&=\frac{1}{\epsilon_{p}^{5}}e^2E_0^2\tau^2\left(\begin{array}{c}
        p_{x}(\epsilon_{p}^{2}-3p_{z}^{2})\\
        p_{y}(\epsilon_{p}^{2}-3p_{z}^{2})\\
        3p_{z}(\epsilon_{p}^{2}-p_{z}^{2})
        \end{array}\right),\label{eq:v_order2}\\
   \bm{a}^{(2)}(\bm{p},\tau)&=\frac{2}{\epsilon_{p}} e E_0 \tau^2 \left(\begin{array}{c}
        p_{y}\\
        -p_x\\
        0 \end{array}\right),\\
    \bm{t}_{1}^{(2)}(\bm{p},\tau)&=-\frac{2m}{\epsilon_{p}} e E_0 \tau^2 \left(\begin{array}{c}
        0\\
        0\\
        1 \end{array}\right).
\end{align}
Other components are zeros.

The third order result is
\begin{widetext}
    \begin{align}
        s^{(3)}(\bm{p},\tau)&=-\frac{m}{\epsilon_{p}^{3}}e E_0 \tau^3 p_{z}\left[\frac{1}{\epsilon_{p}^{4}}e^2 E_0^2 (3\epsilon_{p}^{2}-5p_{z}^{2})+\frac{4}{3}\epsilon_{p}^{2}\right],\\
        a_{0}^{(3)}(\bm{p},\tau)&=-\frac{2}{3}\frac{1}{\epsilon_{p}^{3}} e^2 E_0 B_0 \tau^3 (\epsilon_{p}^{2}+p_{z}^{2}+m^{2}),\label{eq:a0_order3}\\
        \bm{v}^{(3)}(\bm{p},\tau)&=-\frac{1}{\epsilon_{p}^{3}}e E_0 \tau^3 \left(\begin{array}{c}
            p_{x}p_{z}\left[\frac{1}{\epsilon_{p}^{4}}\left(3\epsilon_{p}^{2}-5\bar{p}_{z}^{2}\right)e^2E_0^2+\frac{4}{3}\epsilon_{p}^{2}\right]\\
            p_{y}p_{z}\left[\frac{1}{\epsilon_{p}^{4}}\left(3\epsilon_{p}^{2}-5\bar{p}_{z}^{2}\right)e^2 E_0^2 +\frac{4}{3}\epsilon_{p}^{2}\right]\\
            \frac{1}{\epsilon_{p}^{4}}e^2 E_0^2 \left(6\epsilon_{p}^{2}p_{z}^{2}-5p_{z}^{4}-\epsilon_{p}^{4}\right)+\frac{4}{3}\epsilon_{p}^{2} \left(p_{z}^{2}-\epsilon_{p}^{2}\right)
            \end{array}\right),\\
        \bm{a}^{(3)}(\bm{p},\tau)&=\frac{2}{\epsilon_{p}^{3}} e^2 E_0^2 \tau^3 p_{z}\left(\begin{array}{c}
            -p_{y}\\
            p_x\\
            0 \end{array}\right),\\
        \bm{t}_{1}^{(3)}(\bm{p},\tau)&=\frac{2m}{\epsilon_{p}^{3}} e^2 E_0^2 \tau^3 p_{z}\left(\begin{array}{c}
            0\\
            0\\
            1 \end{array}\right),\\
        \bm{t}_{2}^{(3)}(\bm{p},\tau)&=-\frac{2}{3}\frac{m}{\epsilon_{p}^{3}} e^2 E_0 B_0 \tau^3
            \left(\begin{array}{c}
            p_{x}\\
            p_y\\
            0
            \end{array}\right)\label{eq:t2_order3}.
    \end{align}
    Other components are zeros.

    The fourth order result is
    \begin{align}
        s^{(4)}(\bm{p},\tau) &= -\frac{m}{12\epsilon_{p}^{9}}e^{2} E_0^2 \tau^4 \left[3e^{2} E_0^2 \left(35p_{z}^{4}-30p_{z}^{2}\epsilon_{p}^{2}+3\epsilon_{p}^{4}\right)+4\epsilon_{p}^{6}\left(\epsilon_{p}^{2}-4p_{z}^{2}\right)\right],\\
        s_p^{(4)}(\bm{p},\tau) &= \frac{2m}{3\epsilon_{p}}e^{2}E_0B_0\tau^{4},\label{eq:p_order4}\\
        a_0^{(4)}(\bm{p},\tau) &= \frac{2}{3\epsilon_{p}^{5}}e^{3}E_0^2B_0\tau^4 p_{z}(3m^{2}+3p_{z}^{2}-\epsilon_{p}^{2}),\label{eq:a0_order4}\\
        \bm{v}^{(4)}(\bm{p},\tau) &= \frac{1}{12\epsilon_{p}^{9}}e^{2} E_0^2 \tau^4 \left(\begin{array}{c}
            p_{x}\left[-3e^{2} E_0^2 \left(35p_{z}^{4}-30p_{z}^{2}\epsilon_{p}^{2}+3\epsilon_{p}^{4}\right)+4 \epsilon_{p}^{6}\left(4p_{z}^{2}-\epsilon_{p}^{2}\right)\right]\\
            p_{y}\left[-3e^{2}E_0^2 \left(35p_{z}^{4}-30p_{z}^{2}\epsilon_{p}^{2}+3\epsilon_{p}^{4}\right)+4\epsilon_{p}^{6}\left(4p_{z}^{2}-\epsilon_{p}^{2}\right)\right]\\
            -p_{z}\left[15e^{2}E_0^2 \left(7p_{z}^{4}-10p_{z}^{2}\epsilon_{p}^{2}+3\epsilon_{p}^{4}\right)+16\epsilon{}_{p}\left(\epsilon_{p}^{2}-p_{z}^{2}\right)\right]
            \end{array}\right),\\
        \bm{a}^{(4)}(\bm{p},\tau)&=e E_0 \tau^4 \left[-\frac{1}{\epsilon_{p}^{5}}e^{2} E_0^2 (3p_{z}^{2}-\epsilon_{p}^{2})+\frac{1}{6\epsilon_{p}^{5}}e^{2} B_0^2 (3m^{2}+3p_{z}^{2}+\epsilon_{p}^{2})+\frac{2}{3}\epsilon_{p}\right]\left(\begin{array}{c}
            -p_{y}\\
            p_{x}\\
            0
            \end{array}\right),\\
        \bm{t}_{1}^{(4)}(\bm{p},\tau) &= e E_0 \tau^4 m\left[-\frac{1}{\epsilon_{p}^{5}}e^{2} E_0^2 (3p_{z}^{2}-\epsilon_{p}^{2})+\frac{1}{6\epsilon_{p}^{5}}e^{2} B_0^2 (\tau)(3m^{2}+3p_{z}^{2}-\epsilon_{p}^{2})+\frac{2}{3}\epsilon_{p}\right]\left(\begin{array}{c}
            0\\
            0\\
            1
            \end{array}\right),\\
        \bm{t}_{2}^{(4)}(\bm{p},\tau) &= \frac{2m}{\epsilon_{p}^{5}} e^{3} E_0^2 B_0 \tau^4 p_{z}\left(\begin{array}{c}
            p_{x}\\
            p_{y}\\
            0 \end{array}\right).
    \end{align}
    Other components are zeros.
\end{widetext}

\section{Derivation of the First Equation of Eqs.~(\ref{eq:eqs_chiral}) from Dirac Equation\label{sec:dev-chiral}}

Apart from the derivation based on Dirac-Heisenberg-Wigner formalism given in the main text, in this section, we also provide a derivation of the equation $\partial_{t}s_p+2ma_{0}+2\bm{p}\cdot\bm{t}_{2}=0$ based on the Dirac equation.

As the external fields vanish at $t>\tau$, we would use the Dirac equation for free particles:
\begin{align}
    \partial_{t}\psi(x)=-\gamma^{0}\left(\gamma^{i}\partial_{i}\psi(x)+im\psi(x)\right),\label{eq:Dirac}\\
    \partial_{t}\bar{\psi}(x)=-\left(\partial_{i}\bar{\psi}(x)\gamma^{i}-im\bar{\psi}(x) \right)\gamma^{0}.\label{eq:Dirac-conj}
\end{align}

What we want to compute is the time derivative of the pseudoscalar condensate:
\begin{equation}
    \partial_{t}\langle \bar{\psi} i\gamma_{5} \psi \rangle (t) = \int d^{3}\bm{x}\partial_{t}\bar{\psi}(x)i\gamma_{5}\psi(x)+\int d^{3}\bm{x}\bar{\psi}(x)i\gamma_{5}\partial_{t}\psi(x).\label{eq:pseudo-time-diff}
\end{equation}

After substituting Eqs.~(\ref{eq:Dirac}),(\ref{eq:Dirac-conj}) into Eq.~(\ref{eq:pseudo-time-diff}), and use the integration by parts, we arrive at
\begin{align}
    \partial_{t}\langle \bar{\psi} i\gamma_{5} \psi \rangle (t)=&i\int d^{3}\bm{x}\partial_{i}\bar{\psi}(x)(\gamma_{5}\gamma^{0}\gamma^{i}-\gamma^{i}\gamma^{0}\gamma_{5})\psi(x)\nonumber\\
    &+ m\int d^{3}\bm{x}\bar{\psi}(x)[\gamma_{5},\gamma^{0}]\psi(x).\label{eq:pseudo-time-diff-simp1}
\end{align}

After doing some Dirac algebra using $\{\gamma^\mu,\gamma^\nu\}=2\eta^{\mu\nu}$ ($\eta^{\mu\nu}$ as the Minkovskii metric with $\eta^{00}=+1$), we would have
\begin{align}
    [\gamma_{5},\gamma^{0}]&=-2\gamma^{0}\gamma_{5},\\
    \gamma_{5}\gamma^{0}\gamma^{i}-\gamma^{i}\gamma^{0}\gamma_{5}&=2\epsilon^{ijk}\sigma_{jk}.
\end{align}
Using these relations, Eq.~(\ref{eq:pseudo-time-diff-simp1}) becomes
\begin{align}
    \partial_{t}\langle \bar{\psi} i\gamma_{5} \psi \rangle (t)=&2\int d^{3}\bm{x}i\partial_{i}\bar{\psi}(x)\epsilon^{ijk}\sigma_{jk}\psi(x)\nonumber\\
    &-2m\int d^{3}\bm{x}\bar{\psi}(x)\gamma^{0}\gamma_{5}\psi(x).\label{eq:pseudo-time-diff-simp2}
\end{align}

As the final step, recalling that for the Fourier transformation defined as $f(\bm{x})=\int d^{3}\bm{k}/(2\pi)^{3}\tilde{f}(\bm{k})e^{i\bm{k}\cdot\bm{x}}$, we have
\begin{align}
    \int d^{3}\bm{x}f^{*}(\bm{x})f(\bm{x})=\int\frac{d^{3}\bm{k}}{(2\pi)^{3}}\tilde{f}^{*}(\bm{k})\tilde{f}(\bm{k}).
\end{align}
This would transform Eq.~(\ref{eq:pseudo-time-diff-simp2}) as
\begin{align}
    \partial_{t}\langle \bar{\psi} i\gamma_{5} \psi \rangle (t)=\int\frac{d^{3}\bm{p}}{(2\pi)^{3}}&\left[-2p_i\bar{\psi}(\bm{p},t)\epsilon^{ijk}\sigma_{jk}\psi(\bm{p},t)\right.\nonumber\\
    &\left.-2m\bar{\psi}(\bm{p},t)\gamma^{0}\gamma_{5}\psi(\bm{p},t)\right] .\label{eq:pseudo-time-diff-simpf}
\end{align}
In the same time, from Eq.~(\ref{eq:spatial-averages}), we know
\begin{equation}
    \langle \bar{\psi} i\gamma_{5} \psi \rangle (t) = -V \int \frac{d^3\bm{p}}{(2\pi)^3} s_p(\bm{p},t),
\end{equation}
\begin{equation}
    \int\frac{d^{3}\bm{p}}{(2\pi)^{3}}\bar{\psi}(\bm{p},t)\epsilon^{ijk}\sigma_{jk}\psi(\bm{p},t) = -V \int \frac{d^3\bm{p}}{(2\pi)^3} \left(\bm{t}_2(\bm{p},t)\right)^i,
\end{equation}
\begin{equation}
    \int \frac{d^3\bm{p}}{(2\pi)^3} \bar{\psi}(\bm{p},t)\gamma^{0}\gamma_{5}\psi(\bm{p},t) = -V \int \frac{d^3\bm{p}}{(2\pi)^3} a_{0}(\bm{p},t).
\end{equation}
This shows that Eq.~(\ref{eq:pseudo-time-diff-simpf}) is nothing more than $\partial_{t}s_p+2ma_{0}+2\bm{p}\cdot\bm{t}_{2}=0$.

\bibliographystyle{apsrev4-2}
\bibliography{main.bib}

\end{document}